\documentclass[10pt,journal,compsoc]{IEEEtran}
\usepackage{booktabs,multirow,indentfirst,graphicx,subfig,cite}
\ifCLASSINFOpdf
\else
\fi
\begin{document}
%
\title{HI-Series Algorithms: A Hybrid of Substance Diffusion Algorithm and Collaborative Filtering}
%
%
%
%

\author{Yu Peng, Ya-Hui AN
\IEEEcompsocitemizethanks{\IEEEcompsocthanksitem College of Computer and Cyber Security, Hebei Normal University, Shijiazhuang,
China, 050024.\protect\\
E-mail: anyahui.120@gmail.com}
\thanks{}}

%
%

\markboth{}%
{Shell \MakeLowercase{\textit{et al.}}: HI-Series Algorithms: A Hybrid of Substance Diffusion Algorithm and Collaborative Filtering}

\IEEEtitleabstractindextext{%
\begin{abstract}
 Recommendation systems face the challenge of balancing accuracy and diversity, as traditional collaborative filtering (CF) and network-based diffusion algorithms exhibit complementary limitations. While item-based CF (ItemCF) enhances diversity through item similarity, it compromises accuracy. Conversely, mass diffusion (MD) algorithms prioritize accuracy by favoring popular items but lack diversity. To address this trade-off, we propose the HI-series algorithms, hybrid models integrating ItemCF with diffusion-based approaches (MD, HHP, BHC, BD) through a nonlinear combination controlled by parameter $\epsilon$. This hybridization leverages ItemCF's diversity and MD's accuracy, extending to advanced diffusion models (HI-HHP, HI-BHC, HI-BD) for enhanced performance. Experiments on MovieLens, Netflix, and RYM datasets demonstrate that HI-series algorithms significantly outperform their base counterparts. In sparse data ($20\%$ training), HI-MD achieves a $ 0.8\%$–$4.4\%$ improvement in F1-score over MD while maintaining higher diversity (Diversity@20: 459 vs. 396 on MovieLens). For dense data ($80\%$ training), HI-BD improves F1-score by $2.3\%$–$5.2\%$ compared to BD, with diversity gains up to $18.6\%$. Notably, hybrid models consistently enhance novelty in sparse settings and exhibit robust parameter adaptability. The results validate that strategic hybridization effectively breaks the accuracy-diversity trade-off, offering a flexible framework for optimizing recommendation systems across data sparsity levels.
\end{abstract}

}

\maketitle{}

\IEEEdisplaynontitleabstractindextext

%
\IEEEpeerreviewmaketitle

\ifCLASSOPTIONcompsoc
\IEEEraisesectionheading{\section{Introduction}\label{sec:introduction}}
\else
\section{Introduction}
\label{sec:introduction}
\fi

%
%
%
%
\IEEEPARstart{R}{ecommendation} systems, pivotal in mitigating information overload, leverage user-item interaction histories to model user preferences and predict future engagements. Traditional approaches predominantly fall into three categories: collaborative filtering (CF), network-based diffusion methods, and hybrid models. While CF algorithms, particularly item-based CF (ItemCF), excel in diversifying recommendations by identifying item similarities through co-occurrence patterns, they often compromise accuracy due to over-reliance on associative rules. Conversely, mass diffusion (MD) algorithms, inspired by physical propagation dynamics, prioritize accuracy by favoring popular items, yet lack personalization and diversity.  

The inherent trade-off between accuracy and diversity persists as a critical challenge. ItemCF's strength in diversity is counterbalanced by its suboptimal accuracy, whereas MD's accuracy diminishes its ability to recommend niche items. Recent hybrid models like HHP attempt to balance these aspects but remain constrained by their design frameworks.  

In this study, we propose the HI-series algorithms—HI-MD, HI-HHP, HI-BHC, and HI-BD—a novel hybrid framework integrating ItemCF with diffusion-based methods. By nonlinearly combining item similarity metrics from ItemCF and resource diffusion mechanisms, our approach synergizes the accuracy of diffusion processes with the diversity of CF. Evaluations on MovieLens, Netflix, and RYM datasets demonstrate that HI-series algorithms significantly outperform baseline methods in both sparse (training ratio $(E^T=20\%)$ and dense $(E^T=80\%)$ data scenarios. Specifically, HI-MD improves F1-score by $2.3–5.9\%$ while maintaining competitive diversity, and HI-BD achieves a $12.7\%$ increase in coverage diversity without sacrificing accuracy.  
This work contributes: (1) a systematic hybrid methodology bridging CF and diffusion algorithms, (2) empirical validation across multiple real-world datasets, and (3) insights into parameter tuning for optimizing accuracy-diversity trade-offs. The HI-series framework offers a robust solution for modern recommendation systems, particularly in scenarios requiring balanced performance across metrics.

\hfill 
 
\hfill

\section{Problem Statement}
Recommendation systems establish user interest models based on the interaction history between users and the system, as well as the personal information of the users, to predict potential items that users might purchase or click in the future. With recommendation systems, users can discover items they like without wasting a lot of time searching on the Internet, and it can increase the ratio of cross-selling items. Additionally, by providing users with items that fully meet their requirements, it will increase the trust users have in the system and prompt them to make purchases. Currently, the most extensive recommendation algorithms, which are purely based on historical interaction triple data (user, item, action) without introducing other information, allowing researchers to conduct more pure model research, can be divided into three types: collaborative filtering algorithms\cite{1,2,3,4}, Network Structure-based Diffusive Recommendation Algorithm\cite{5,6,7,8}, Hybrid Algorithm\cite{17,18}.

Collaborative filtering methods produce recommendations based on usage patterns\cite{19}, and collaborative filtering based on user similarity or item similarity can be divided into two main types: User-based Collaborative Filtering (UserCF)\cite{9,10} and Item-based Collaborative Filtering (ItemCF)\cite{11,12,13,14}\cite{24}.

The main idea of user-based collaborative filtering is to recommend items that have been chosen by similar users of the target user but have not yet been selected by the target user, while item-based collaborative filtering mainly focuses on items similar to those previously chosen by the target user\cite{13}. Therefore, the method of calculating the similarity between users or items becomes the most important part in the collaborative filtering process. In order to obtain more accurate similarity, it requires more information.Collaborative filtering provides benefits to most online stores in recommending products based on users' ratings of similarity measurement, but its usage also raises doubts among researchers regarding its effectiveness in handling ratings with a limited number of users or no rating record from users\cite{22}, and there are significant flaws in user-based measurement\cite{23}. For example, on the e-commerce platform Taobao, user behavior data such as browsing, clicking, adding to the shopping cart, and purchasing are available. As for the algorithm itself, UserCF is based on the similarity between users. The "similarity" of users similar to the target user is calculated based on the similarity of items purchased by the users. As a result, users will be recommended similar items, and the recommendation results among users lack personalization (see the table\ref{tab:motivation}).

On the contrary, ItemCF is based on the similarity between items, and the similarity between items is often measured by common users. Therefore, the similarity between items is more similar to the traditional recommendation algorithm---association rules. For example, diapers and beer are common choices for many users. According to the idea of ItemCF, diapers and beer are similar, but they are actually products of completely different categories. So ItemCF can achieve diversification of the recommended list for a single user, but the accuracy of the recommendation results based on this association rule mode is not high (see the table\ref{tab:motivation}).

In recent years, with the development of recommendation technology, Pulkit Dwivedi\cite{25} and others propose an item-based CF method where they initially analyze the user's item rating pattern and then they establish the connections between several items. Zhou\cite{15} and others have constructed a simple model to solve the recommendation problem from the perspective of physical propagation, referred to as the Mass Diffusion Algorithm (MD). This method utilizes network structure information and has higher accuracy and diversity than UserCF (see table\ref{tab:motivation}). However, in order to ensure accuracy, this method is not as flexible as the ItemCF algorithm in terms of overall system diversity and personalization among users. Similarly, by simulating the physical propagation process, the Heat Conduction Algorithm (HC) was also introduced by Zhou and other\cite{16}. The HC algorithm can achieve higher diversity than Mass Diffusion, even much higher than ItemCF, but the accuracy of this algorithm is extremely low, basically making it unable to be applied. Therefore, this paper simply compares the performance of this algorithm with the other three basic algorithms without conducting in-depth research. Subsequently, researchers found that Mass Diffusion tends to recommend popular items, while Heat Conduction tends to recommend unpopular ones. In order to balance the trade-off between being able to recommend popular items with high accuracy and being able to recommend unpopular items with high diversity, a hybrid model, HHP, was proposed, which effectively solves the problem of the accuracy-diversity trade-off.

We are inspired by the fact that since ItemCF has high diversity and its accuracy is within an acceptable range (while the accuracy of HC is basically unacceptable), and MD has high accuracy and its diversity is also within an acceptable range, we can mix these two algorithms and believe that we can obtain better accuracy and diversity performance than HHP. Based on this, we propose the hybrid algorithm HI-MD of ItemCF and MD, and the mixing idea can also be further applied in a series of diffusion-like algorithms HHP, BHC, and BD. The new algorithms are named as HI-HHP, HI-BHC, and HI-BD in turn.

\section{An Empirical Analysis of the Advantages and Disadvantages of Existing Algorithms}

To test the effectiveness of the algorithms, We used three datasets: MovieLens, Netflix, and RYM. In the experiments, each dataset was randomly divided into two subsets based on the edges: $80\%$ was used as the training set and the remaining $20\%$ as the test set. The MovieLens dataset constructed in this way was named MovieLens ($E^T$=80). And the sparse dataset with $20\%$ used as the training set and the remaining $80\%$ as the test set was named MovieLens ($E^T$=20). The other two datasets were also named following a similar rule. This chapter mainly studies the accuracy and diversity of hybrid algorithms. Therefore, the accuracy indicators selected were Precision@K, Recall@K, and F1-Score@K\cite{26}, while the diversity indicators chosen were system global coverage diversity Diversity-in-top-K and inter-user recommendation list diversity HD@K.

So far, collaborative filtering has become the most widely used mainstream recommendation technology. UserCF and ItemCF are two main branches of collaborative filtering algorithms\cite{19,20,21}. The main idea of ItemCF is to recommend items similar to those that users like. For two items $\alpha $ and $\beta$, $s_{\alpha \beta}$ represents the similarity between the two items. On implicit feedback data, the cosine similarity between two items is:
\vspace{-5mm}
\begin{center}
	\begin{equation}
	\label{equ:itemcf}
	s^{ItemCF}_{\alpha \beta} = \frac{\sum_{l=1}^{m}{a_{l\alpha} a_{l\beta}}}{\sqrt{k_{\alpha} k_{\beta}}}
	\end{equation}
\end{center}

Its advantage lies in the ability to consider the antagonism of items based on their degree of pairing. For example, suppose two pairs of items have the same number of common neighbors, but one pair consists of popular items, while the other pair consists of unpopular items. In this case, the similarity between the two pairs of items should be treated differently. ItemCF based on cosine similarity can precisely distinguish these pairs of items. However, this method has its flaws. In the scale-free networks of the real world, most items are not very popular. When we use ItemCF, there is a tendency to recommend unpopular items to users. Therefore, ItemCF can achieve high diversity, but its accuracy is not high.

In recent years, Zhou and others made recommendations by simulating the process of material diffusion in physics, which is an algorithm that focuses on accuracy. Through two-step diffusion, the resources of the items already collected by the target user are allocated to the items that the target user has not yet chosen, and the similarity between two items can be expressed as:

\vspace{-5mm}
\begin{center}
	\begin{equation}
	\label{equ5-6}
	s^{MD}_{\alpha\beta} = \frac{1}{k_{\beta}}\sum_{l=1}^{m}{\frac{a_{l\alpha}a_{l\beta}}{k_{l}}}
	\end{equation}
\end{center}

Among them, $s^{MD}_{\alpha\beta} \neq s^{MD}_{\beta\alpha}$. Obviously, the more popular an item is, the greater the probability that it will obtain more resources. That is to say, the material diffusion algorithm tends to recommend popular items. We found that the similarity between most pairs of items is low, especially among unpopular items, which further verifies our inference: resources are concentrated on items with greater popularity, so the material diffusion tends to popular items and will ultimately achieve relatively accurate recommendation results.

	\begin{table*}[!ht]
	\centering
	\caption{Comparison of the effects of ItemCF, UserCF, MD, and HC}
	\label{tab:motivation}
	\begin{tabular}{cccccccc}
		\hline
		&&\multicolumn{3}{c}{Accuracy}&\multicolumn{2}{c}{Diversity}&Novelty\\ \cmidrule(lr){3-5}  \cmidrule(lr){6-7} \cmidrule(lr){8-8}
		Data set&Algorithm&Precision&Recall&F1-Score&Diversity-&HD&Novelty\\
		&&@20&@20&@20&in-top-20&@20&@20\\
		\hline
		\multirow{3}{*}{MovieLens}
		&ItemCF&0.378&0.112&0.173&\textbf{1102}&\textbf{0.854} &4.85\\
		&UserCF&0.485&0.175&0.257&224&0.546 &3.68\\
		($E^T$=20)&MD&\textbf{0.495}&\textbf{0.179}&\textbf{0.263}&396&0.650&3.81 \\
		&HC&0.030&0.011&0.016&722&\textbf{0.912}&\textbf{9.66}\\
		\hline
		\multirow{3}{*}{Netflix}
		&ItemCF&0.326&0.160&0.214&\textbf{3772}&\textbf{0.724}&5.16 \\
		&UserCF&0.352&0.192&0.248&708&0.398 & 3.94\\
		($E^T$=20)&MD&\textbf{0.360}&\textbf{0.197}&\textbf{0.254}&1708&0.523 &4.10 \\
		&HC&0.001&0.001&0.001&2755&\textbf{0.946}&\textbf{12.95}\\
		\hline
		\multirow{3}{*}{RYM}
		&ItemCF&0.146&0.210&0.173&\textbf{4790}&\textbf{0.976}&10.36 \\
		&UserCF&0.200&0.308&0.243&2920&0.693 &7.01\\
		($E^T$=20)&MD&\textbf{0.209}&\textbf{0.312}&\textbf{0.251}&3520&0.753 &7.30\\
		&HC&0.019&0.049&0.028&4497&\textbf{0.974}&\textbf{13.37}\\
		\hline
		\hline			
		\multirow{3}{*}{MovieLens}
		&ItemCF&\textbf{0.235}&\textbf{0.291}&\textbf{0.260}&\textbf{272}&\textbf{0.792} & 2.05\\
		&UserCF&0.210&0.259&0.232&128&0.638 &1.73\\
		($E^T$=80)&MD&0.224&0.289&0.253&162&0.695&1.82 \\
		&HC&0.049&0.046&0.047&\textbf{864}&\textbf{0.891}&\textbf{7.29}\\
		\hline		
		\multirow{3}{*}{Netflix}
		&ItemCF&\textbf{0.146}&0.264&\textbf{0.188}&\textbf{831}&\textbf{0.618}&2.27 \\
		&UserCF&0.133&0.251&0.174&139&0.475 &1.97\\
		($E^T$=80)&MD&0.140&\textbf{0.269}&0.184&262&0.523&2.00 \\
		&HC&0.0002&0.0005&0.0003&\textbf{1262}&\textbf{0.838}&\textbf{12.93}\\
		\hline						
		\multirow{3}{*}{RYM}
		&ItemCF&0.097&0.458&0.160&\textbf{4368}&\textbf{0.940} & 6.24\\
		&UserCF&0.085&0.414&0.140&1881&0.669 & 4.91\\
		($E^T$=80)&MD&\textbf{0.102}&\textbf{0.485}&\textbf{0.169}&2704&0.709 &5.07\\
		&HC&0.067&0.340&0.112&\textbf{4611}&\textbf{0.945}&\textbf{10.16}\\
		\hline							
	\end{tabular}
\end{table*}
As shown in Table \ref{tab:motivation}, it presents the results of accuracy, diversity, and novelty of the ItemCF, UserCF, MD, and HC algorithms on sparse data ($E^T$=20) and relatively dense datasets ($E^T$=80)(with a recommendation list length of 20).It can be seen that on the $E^T$=20 data, the material diffusion algorithm has the best accuracy among the three datasets, while ItemCF can expose the most items (Diversity-in-top-20). However, in terms of the diversity and novelty among the recommendation lists of each user, the heat conduction algorithm performs the best. But the accuracy of the heat conduction algorithm is too low, making it basically unusable in real life. Therefore, after removing the heat conduction algorithm, both the diversity and novelty among the recommendation lists of ItemCF users are the best.

On the denser datasets, except for the RYM dataset, ItemCF performs better in terms of both accuracy and diversity, followed by material diffusion. Moreover, on the RYM ($E^T$=20), the accuracy of material diffusion is good, but its diversity is not as good as that of ItemCF. Overall, whether on sparse or dense data, the diversity of ItemCF is relatively stable. The accuracy of material diffusion is relatively stable on sparse data, but on dense datasets, its accuracy is slightly lower than, equal to, or slightly higher than that of ItemCF, but the overall accuracy is still good. The accuracy of the heat conduction algorithm is extremely low, and the accuracy and diversity of UserCF are not significant. In conclusion, we have chosen to mix the accuracy of material diffusion with the diversity and novelty of collaborative filtering based on item similarity. It can be estimated that the mixed algorithm will perform well on sparse data; it will also improve the accuracy, diversity, and novelty on dense datasets, but the effect may not be as noticeable as on sparse data. In the next section, let's look at the model function of the mixed algorithm and its specific performance when applied to the best class diffusion algorithm so far.

\section{HI-series algorithm}
\subsection{Algorithm description}

In the choice between linear and non-linear mixing, one is to maintain consistency with the non-linear mixing of the HHP hybrid algorithm, and the other is because ItemCF and MD are not two independent algorithms, so we chose a non-linear mixing method. The first algorithm to be introduced is the ItemCF and MD hybrid algorithm (HI-MD). The similarity between two items $\alpha$ and $\beta$ is affected by MD and is asymmetric, so the similarity between $\alpha$ and an item $\beta$ that the target user has chosen is:
\begin{eqnarray}
\label{equ:hi-md}
\begin{array}{lll}
s^{HI-MD}_{\alpha \beta} & = & (s^{ItemCF}_{\alpha \beta})^{\epsilon} \cdot (s^{MD}_{\alpha\beta})^{1-\epsilon} \\

& = & (\frac{\sum_{l=1}^{m}{a_{l\alpha}a_{l\beta}}}{\sqrt{k_{\alpha} k_{\beta}}})^{\epsilon} \cdot (\frac{1}{k_{\beta}} \sum_{l=1}^{m}{\frac{a_{l\alpha}a_{l\beta}}{k_{l}}})^{1-\epsilon}
\end{array}
\end{eqnarray}

Obviously, when $\epsilon=0$, the HI-MD algorithm degenerates into the MD algorithm, and when $\epsilon=1$, the HI-MD algorithm degenerates into the ItemCF algorithm.

To verify the scalability of the hybrid algorithm, we apply this hybrid idea to HHP, BHC, and BD, which are currently the best in terms of diffusion algorithm performance and have fewer parameters. HHP is a non-linear mixture of MD and HC models, which attempts to solve the problem that accuracy and diversity (or novelty) replace each other, while improving the accuracy, diversity, and novelty of the recommendation system. BHC is a biased heat conduction model that attempts to reduce the drawback that unpopular items absorb more resources than popular items. This method compensates for the resources absorbed by popular items in the last step of resource propagation in the heat conduction algorithm. The BD algorithm studies the balance between MD and HC when mixed, and achieves a balance between the two algorithms through the adjustment of a parameter. Based on the same complexity as HHP, it greatly improves the accuracy and diversity of the system. The three algorithms mixed with ItemCF are called HI-HHP, HI-BHC, and HI-BD respectively, and the similarities are written as follows:

\vspace{-5mm}
\begin{center}
	\begin{equation}
	\label{equ:hihhp}
	s^{HI-HHP}_{\alpha\beta} = (\frac{\sum_{l=1}^{m}{a_{l\alpha}a_{l\beta}}}{\sqrt{k_{\alpha} k_{\beta}}})^{\epsilon} \cdot (\frac{1}{k_{\alpha}^{1-\lambda}k_{\beta}^{\lambda}}\sum_{l=1}^{m}{\frac{a_{l\alpha}a_{l\beta}}{k_{l}}})^{1-\epsilon}
	\end{equation}
\end{center}
\vspace{-5mm}
\begin{center}
	\begin{equation}
	\label{equ:hibhc}
	s^{HI-BHC}_{\alpha\beta} = (\frac{\sum_{l=1}^{m}{a_{l\alpha}a_{l\beta}}}{\sqrt{k_{\alpha} k_{\beta}}})^{\epsilon} \cdot(\frac{1}{k_{\alpha}^{\lambda}}\sum_{l=1}^{m}{\frac{a_{l\alpha}a_{l\beta}}{k_{l}}})^{1-\epsilon}
	\end{equation}
\end{center}
\vspace{-5mm}
\begin{center}
	\begin{equation}
	\label{equ:hibd}
	s^{HI-BD}_{\alpha\beta} = (\frac{\sum_{l=1}^{m}{a_{l\alpha}a_{l\beta}}}{\sqrt{k_{\alpha} k_{\beta}}})^{\epsilon} \cdot(\frac{1}{(k_{\alpha}k_{\beta})^{\lambda}}\sum_{l=1}^{m}{\frac{a_{l\alpha}a_{l\beta}}{k_{l}}})^{1-\epsilon}
	\end{equation}
\end{center}

When $\epsilon=0$ , the HI-series algorithm degenerates into the HHP, BHC, and BD algorithms respectively. When $\epsilon=1$, the HI-series algorithm degenerates into the ItemCF algorithm.

\subsection{Recommendation performance improvement}

This section first shows the comparison of the effect of each mixed algorithm and the original algorithm on sparse datasets ($E^T$=20) and dense datasets ($E^T$=80)respectively. In the selection of the optimal parameters, if there is only one parameter $\epsilon$, the $\epsilon$ that makes the overall accuracy F1-Score@20 the best is taken as $\epsilon$\_{opt}; If there are two parameters, one is the mixing adjustment parameter $\epsilon$ and the other is the original algorithm's own adjustment parameter $\lambda$, then we will traverse the two parameters and take the $\epsilon$ and $\lambda$ that make the overall accuracy F1-Score@20 the best as $\epsilon$\_{opt} and $\lambda$\_{opt} respectively. The experimental results are shown in Table \ref{tab:finalresults-} and Table \ref{tab:final results2} respectively.

{\bf (1)Comparison between mixed algorithm and original algorithm on dense dataset.}

        \begin{table*}[!h] 
        \scriptsize
	\centering
	\caption{Experimental results of the mixed algorithm and the corresponding original algorithm on $E^T$=80}
	\label{tab:finalresults-}
	\begin{tabular}{ccccccccc}
		\hline
		
		\multirow{3}{*}{Data set}&\multirow{3}{*}{Algorithm}&Optimal parameters&\multicolumn{3}{c}{Accuracy}&\multicolumn{2}{c}{Diversity}&Novelty\\ \cmidrule(r){3-3}  \cmidrule(r){4-6} \cmidrule(r){7-8} \cmidrule(r){9-9}
		&&$\lambda$\_{opt}/&Precision&Recall&F1-Score&Diveristy&HD&Novelty\\
		&&$\epsilon$\_{opt}&@20&@20&@20&@20&@20&@20\\\hline
		\multirow{9}{*}{MovieLens}&MD&na/na&0.224&0.289&0.253&162&0.695&1.82 \\
		&ItemCF&na/na&0.235&0.291&0.260&\textbf{272}&\textbf{0.792} &\textbf{2.05}\\
		&HI-MD&na/0.8&\textbf{0.243}&\textbf{0.300}&\textbf{0.268}&233&0.771 &1.95\\
		\cline{2-9}
		&HHP&0.7/na&0.258&0.324&0.287&379&0.826 &2.15\\
		&HI-HHP&0.8/0.4&\textbf{0.269}&\textbf{0.337}&\textbf{0.299}&\textbf{388}&\textbf{0.850}&\textbf{2.16} \\
		\cline{2-9}
		($E^T$=80)&BHC&0.7/na&0.250&0.318&0.280&336&0.809&2.09 \\
		&HI-BHC&0.9/0.5&\textbf{0.266}&\textbf{0.336}&\textbf{0.297}&\textbf{421}&\textbf{0.860}&\textbf{2.21} \\
		\cline{2-9}
		&BD&0.7/na&0.267&0.327&0.294&518&0.859&\textbf{2.30} \\
		&HI-BD&0.8/0.4&\textbf{0.272}&\textbf{0.340}&\textbf{0.301}&\textbf{519}&\textbf{0.872}&2.30\\\hline
		
		\multirow{9}{*}{Netflix}&MD&na/na&0.140&0.269&0.184&262&0.563&2.00 \\
		&ItemCF&na/na&0.146&0.264&0.188&\textbf{831}&\textbf{0.618}&\textbf{2.27} \\
		&HI-MD&na/0.8&\textbf{0.152}&\textbf{0.277}&\textbf{0.196}&494&0.581&2.10 \\
		\cline{2-9}
		&HHP&0.8/na&0.161&0.299&0.209&1809&0.734& \textbf{3.17}\\
		&HI-HHP&1/0.4&\textbf{0.178}&\textbf{0.327}&\textbf{0.230}&\textbf{2217}&\textbf{0.794}&2.63 \\
		\cline{2-9}
		($E^T$=80)&BHC&0.8/na&0.156&0.293&0.204&1454&0.711& \textbf{3.02}\\
		&HI-BHC&1/0.4&\textbf{0.178}&\textbf{0.327}&\textbf{0.230}&\textbf{2217}&\textbf{0.794}& 2.63\\
		\cline{2-9}
		&BD&0.7/na&0.164&0.306&0.214&\textbf{3287}&0.759&\textbf{3.32} \\
		&HI-BD&0.8/0.4&\textbf{0.177}&\textbf{0.324}&\textbf{0.229}&3254&\textbf{0.774}&2.84 \\\hline
		
		\multirow{9}{*}{RYM}&MD&na/na&0.102&0.485&0.169&2704&0.709&5.07 \\
		&ItemCF&na/na&0.097&0.458&0.160&\textbf{4368}&\textbf{0.940}&\textbf{6.24} \\
		&HI-MD&na/0.6&\textbf{0.105}&\textbf{0.493}&\textbf{0.173}&3359&0.811&5.40 \\
		\cline{2-9}
		&HHP&0.6/na&0.113&0.502&0.184&4536&0.907 &6.14\\
		&HI-HHP&0.7/0.3&\textbf{0.116}&\textbf{0.503}&\textbf{0.189}&\textbf{4756}&\textbf{0.932}&\textbf{6.29} \\\cline{2-9}
		($E^T$=80)&BHC&0.8/na&0.101&0.451&0.165&4746&0.926&\textbf{6.91} \\
		&HI-BHC&0.8/0.5&\textbf{0.113}&\textbf{0.492}&\textbf{0.184}&\textbf{4836}&\textbf{0.938}&6.30 \\
		\cline{2-9}
		&BD&0.6/na&0.118&\textbf{0.512}&0.192&4650&0.926&6.35 \\
		&HI-BD&0.7/0.2&\textbf{0.119}&0.509&\textbf{0.193}&\textbf{4915}&\textbf{0.957}&\textbf{6.75} \\\hline
	\end{tabular} 
\end{table*}

{\bf (2) Comparison between mixed algorithm and original algorithm on sparse dataset.}
        \begin{table*}[!ht] \scriptsize
	\centering
	\caption{Experimental results of the mixed algorithm and the corresponding original algorithm on $E^T$=20}
	\label{tab:final results2}
	\begin{tabular}{ccccccccc}
		\hline
		\multirow{3}{*}{Data set}&\multirow{3}{*}{Algorithm}&Optimal parameters&\multicolumn{3}{c}{Accuracy}&\multicolumn{2}{c}{Diversity}&Novelty\\ \cmidrule(r){3-3}  \cmidrule(r){4-6} \cmidrule(r){7-8} \cmidrule(r){9-9}
		&&$\lambda$\_{opt}/&Precision&Recall&F1-Score&Diveristy&HD&Novelty\\
		&&$\epsilon$\_{opt}&@20&@20&@20&@20&@20&@20\\\hline
		
		\multirow{9}{*}{MovieLens}&MD&na/na&0.495&0.179&0.263&396&0.650&3.81 \\
		&ItemCF&na/na&0.378&0.112&0.173&\textbf{1102}&\textbf{0.854}&\textbf{4.85}\\
		&HI-MD&na/0.2&\textbf{0.500}&\textbf{0.180}&\textbf{0.265}&407&0.667&3.84 \\\cline{2-9}
		&HHP&0.1/na&0.497&0.179&0.264&425&0.676&3.88 \\
		&HI-HHP&0.1/0.2&\textbf{0.501}&\textbf{0.180}&\textbf{0.265}&\textbf{459}&\textbf{0.692}&\textbf{3.89} \\\cline{2-9}
		($E^T$=20)&BHC&0.1/na&0.492&0.180&0.264&293&0.621&3.79 \\
		&HI-BHC&0.1/0.3&\textbf{0.499}&\textbf{0.181}&\textbf{0.266}&\textbf{339}&\textbf{0.657}&\textbf{3.82} \\\cline{2-9}
		&BD&0.1/na&0.494&0.181&0.265&296&0.622&3.79 \\
		&HI-BD&0.1/0.2&\textbf{0.499}&\textbf{0.182}&\textbf{0.266}&\textbf{325}&\textbf{0.646}& \textbf{3.81}\\\hline
		
		\multirow{9}{*}{Netflix}&MD&na/na&0.360&0.197&0.254&1708&0.523&4.10 \\
		&ItemCF&na/na&0.326&0.160&0.214&\textbf{3772}&\textbf{0.724}& \textbf{5.16}\\
		&HI-MD&na/0.4&\textbf{0.367}&\textbf{0.198}&\textbf{0.257}&2502&0.532&4.16 \\\cline{2-9}
		&HHP&0.3/na&0.369&0.202&0.261&3038&0.586&4.36 \\
		&HI-HHP&0.4/0.3&\textbf{0.375}&\textbf{0.202}&\textbf{0.262}&\textbf{3658}&\textbf{0.628}&\textbf{4.55} \\\cline{2-9}
		($E^T$=20)&BHC&0.3/na&0.362&0.202&0.259&2244&0.532&4.15 \\
		&HI-BHC&0.4/0.4&\textbf{0.373}&\textbf{0.202}&\textbf{0.262}&\textbf{3169}&\textbf{0.593}&\textbf{4.31} \\\cline{2-9}
		&BD&0.3/na&0.367&0.203&0.262&2368&0.540&4.17 \\
		&HI-BD&0.4/0.3&\textbf{0.375}&\textbf{0.203}&\textbf{0.264}&\textbf{3354}&\textbf{0.607}&\textbf{4.40} \\\hline

		\multirow{9}{*}{RYM}&MD&na/na&0.209&0.312&0.251&3520&0.753& 7.30\\
		&ItemCF&na/na&0.146&0.210&0.173&\textbf{4790}&\textbf{0.976}& \textbf{10.36}\\
		&HI-MD&na/0.5&\textbf{0.221}&\textbf{0.319}&\textbf{0.262}&4616&0.847 &7.75\\\cline{2-9}
		&HHP&0.2/na&0.216&0.312&0.255&4265&0.836& 7.75\\
		&HI-HHP&0.1/0.4&\textbf{0.223}&\textbf{0.319}&\textbf{0.262}&\textbf{4587}&\textbf{0.852}&\textbf{7.76} \\\cline{2-9}
		($E^T$=20)&BHC&0.2/na&0.201&0.302&0.241&4121&0.809&7.55 \\
		&HI-BHC&0.1/0.5&\textbf{0.218}&\textbf{0.314}&\textbf{0.257}&\textbf{4596}&\textbf{0.849}&\textbf{7.73} \\\cline{2-9}
		&BD&0.3/na&0.212&0.304&0.250&4368&0.860 &7.89\\
		&HI-BD&0.2/0.5&\textbf{0.221}&\textbf{0.312}&\textbf{0.259}&\textbf{4720}&\textbf{0.879}&\textbf{7.96} \\\hline
	\end{tabular} 
\end{table*}

It is better to analyze the two tables together. From the two tables, we can see:

1. Whether on sparse datasets or dense datasets, the trends of advantages and disadvantages of each algorithm are similar.

2. Overall, except for HI-MD, other HI-hybrid algorithms perform better than the original algorithms in both accuracy and diversity. That is, HI-HHP is better than HHP in both accuracy and diversity, HI-BHC is better than BHC in both accuracy and diversity, and HI-BD is better than BD in both accuracy and diversity.

3. The accuracy of HI-MD is better than that of MD and ItemCF. In terms of diversity, HI-MD is better than MD but not as good as ItemCF. Although the diversity of the ItemCF algorithm is better than that of HI-MD, it is far less than that of HHP, BHC or BD algorithms, and even less than that of HI-HHP, HI-BHC, and HI-BD.

4. On sparse data, the novelty of hybrid algorithms is higher than that of the original algorithms. However, on dense datasets, the novelty is unstable on different types of datasets and different algorithms. This is mainly because when hybrid algorithms improve accuracy more on dense datasets (recommending more popular items), they will lose certain novelty.

From the results, whether on sparse datasets or dense datasets, if an algorithm without parameters is to be chosen, ItemCF is better than MD in terms of accuracy, diversity, and novelty; if an algorithm with one parameter can be chosen, the BD algorithm is the best choice in terms of accuracy and diversity; if an algorithm with two parameters can be chosen, HI-BD is the best in terms of accuracy and diversity. In terms of novelty, for denser data, the hybrid algorithm with two parameters has the best accuracy, diversity, and novelty.

\subsection{Changes in algorithm performance with the variation of recommendation parameters}
\begin{figure*}[!ht]
	\centering	
	
	\subfloat[]{
		\label{fig:1-3}
		\includegraphics[width=0.7\columnwidth]{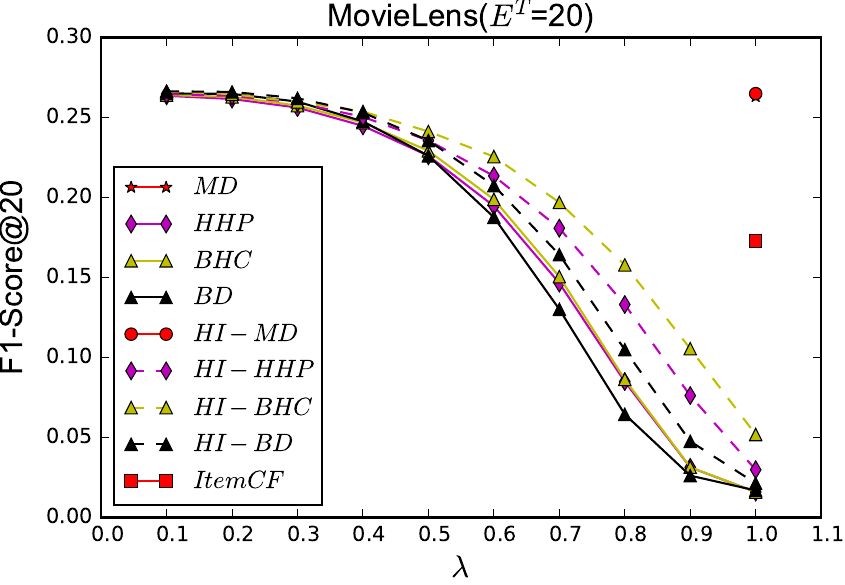}
		\includegraphics[width=0.7\columnwidth]{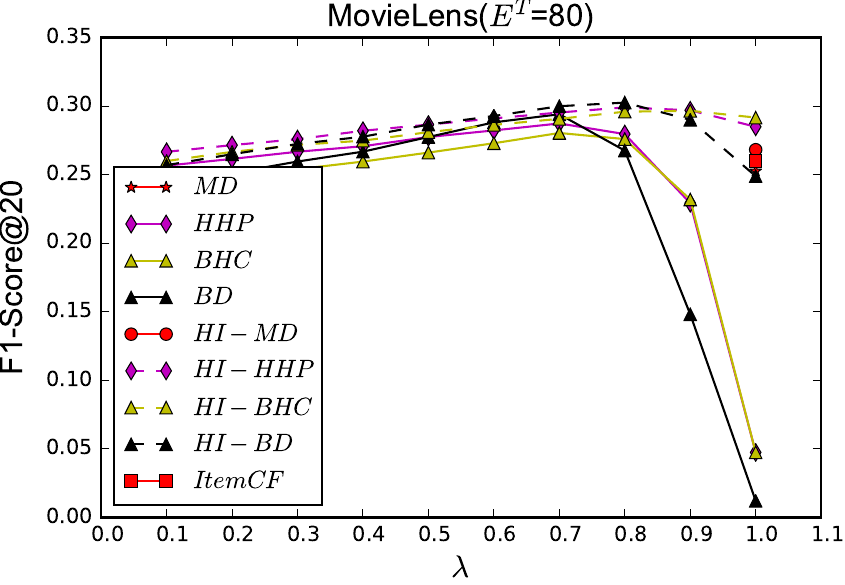}}
        \\
        \quad
	
	\subfloat[]{
		\label{fig:1-4}
		\includegraphics[width=0.7\columnwidth]{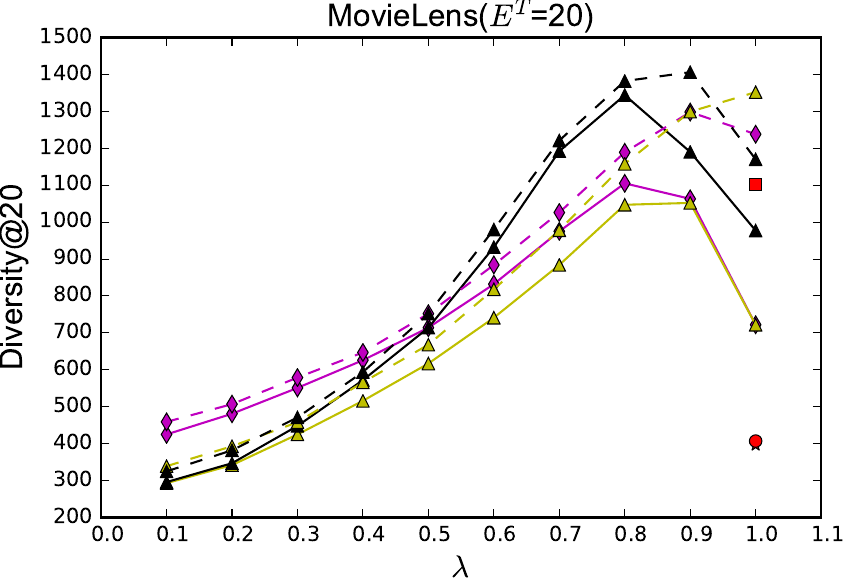}
		\includegraphics[width=0.7\columnwidth]{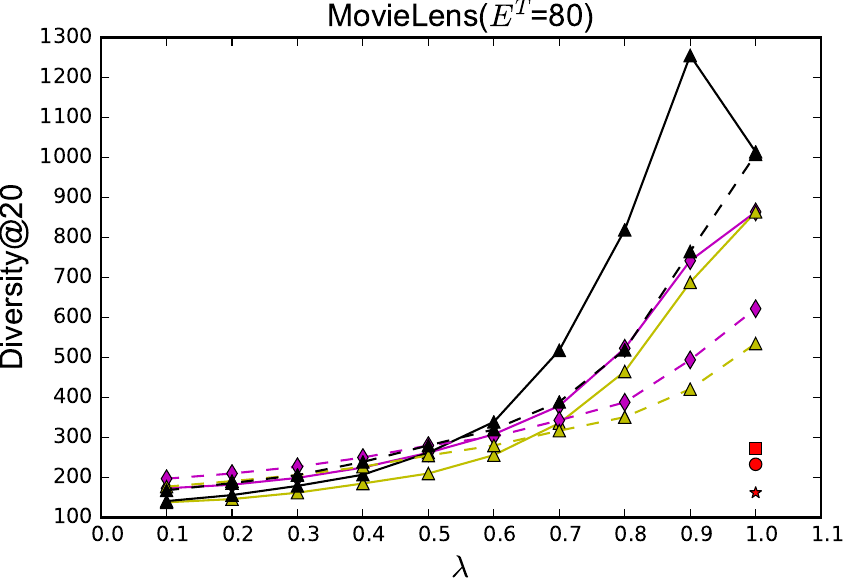}}\\

	
	\subfloat[]{
		\label{fig:1-6}
		\includegraphics[width=0.7\columnwidth]{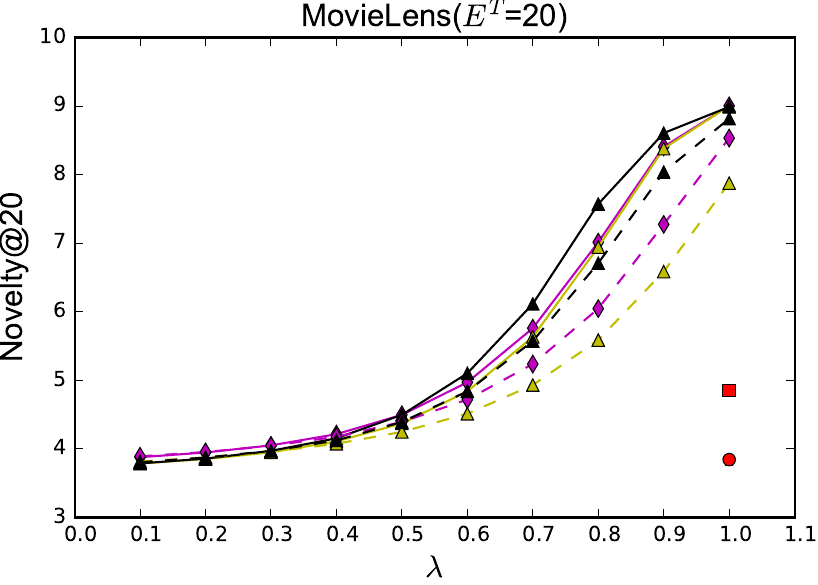}
		\includegraphics[width=0.7\columnwidth]{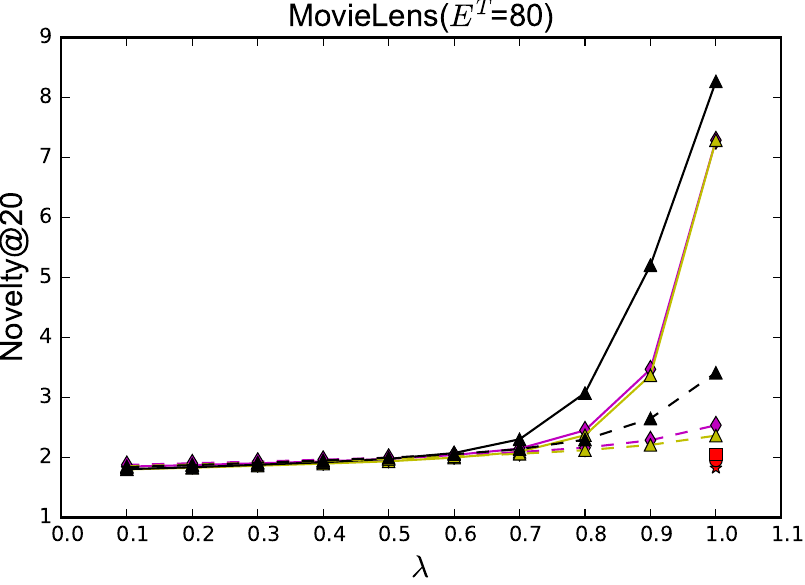}}\\
	\captionsetup{justification=centering}
	\caption{Changes of F1-Score@20, Diversity@20 and Novelty@20 with 
 $\lambda$ on MovieLens.\\(a) F1-Score@20; (b) Diversity@20; (c) Novelty@20.}
	\label{fig:m}
\end{figure*}

Although the performance of accuracy, diversity, and novel. Although the performance of accuracy, diversity, and novel for each algorithm can be clearly and novelty under the optimal parameters for each algorithm can be clearly seen from Table \ref{tab:final results2}, it is not possible to display the trends in which the accuracy, diversity, and novelty of algorithms with $\lambda$ parameters change with respect to $\epsilon$. It is also not possible to show how the accuracy, diversity, and novelty change as the hybrid algorithm transitions from MD having the highest proportion to ItemCF having the highest proportion as $\epsilon$ varies from 0 to 1. Therefore, in the following sections, we will represent the changes in accuracy, diversity, and novelty of various parameterized algorithms with respect to $\lambda$ when $\epsilon$ is at its optimal value, and the changes in accuracy, diversity, and novelty of various hybrid algorithms with respect to $\epsilon$ when $\lambda$ is at its optimal value, using smooth curves for both sparse data ($E^T=20$) and dense datasets ($E^T=80$). To reduce the length, we will only retain F1-Score@20 for accuracy, Diversity@20 for diversity, and Novelty@20 for novelty.
\begin{figure*}[!h]
	\centering	
	
	\subfloat[]{
		\label{fig:2-3}
		\includegraphics[width=0.7\columnwidth]{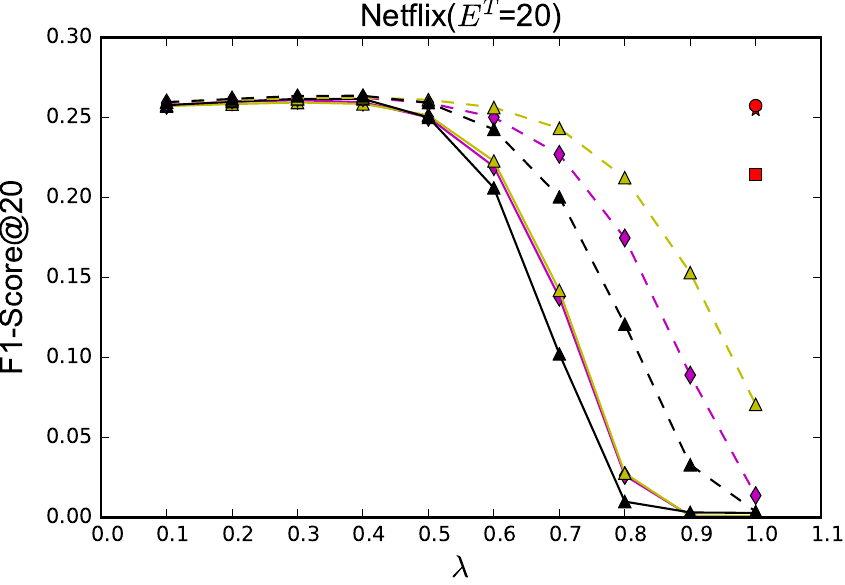}
		\includegraphics[width=0.7\columnwidth]{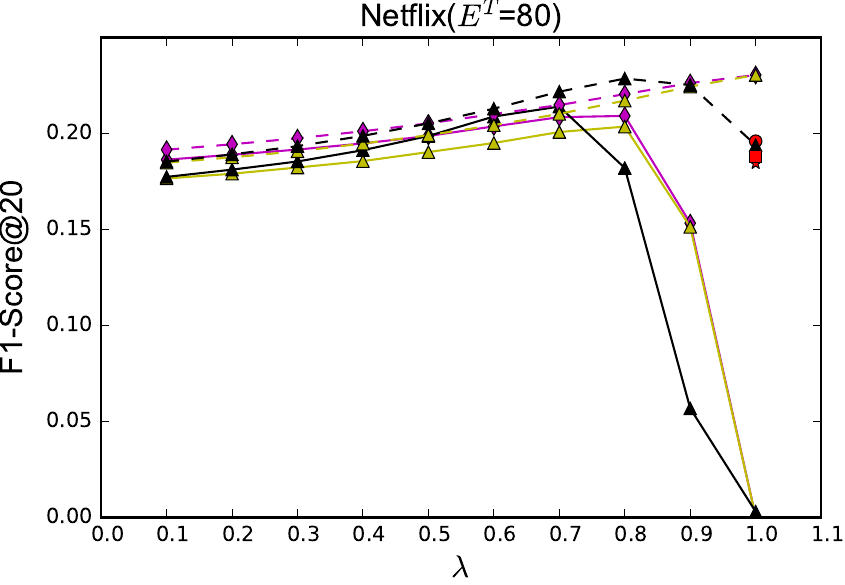}}\\
	\subfloat[]{
		\label{fig:2-4}
		\includegraphics[width=0.7\columnwidth]{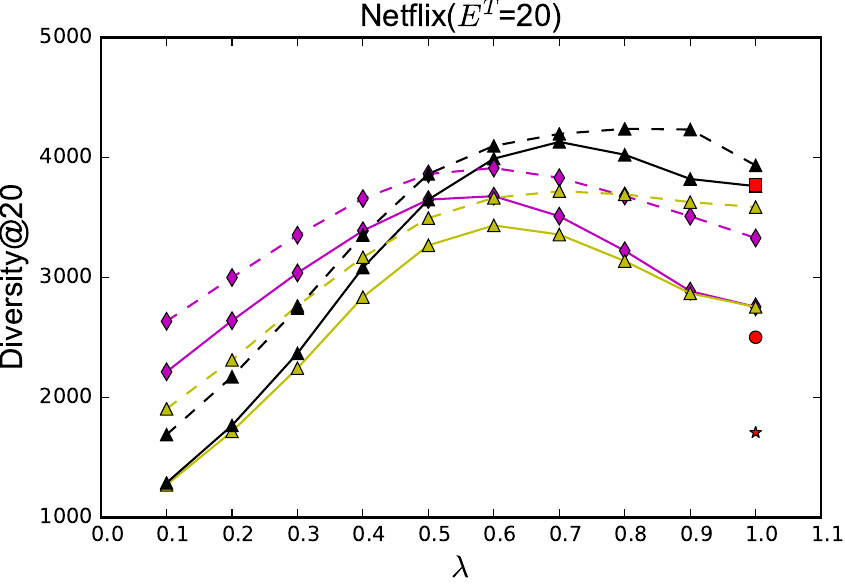}
		\includegraphics[width=0.7\columnwidth]{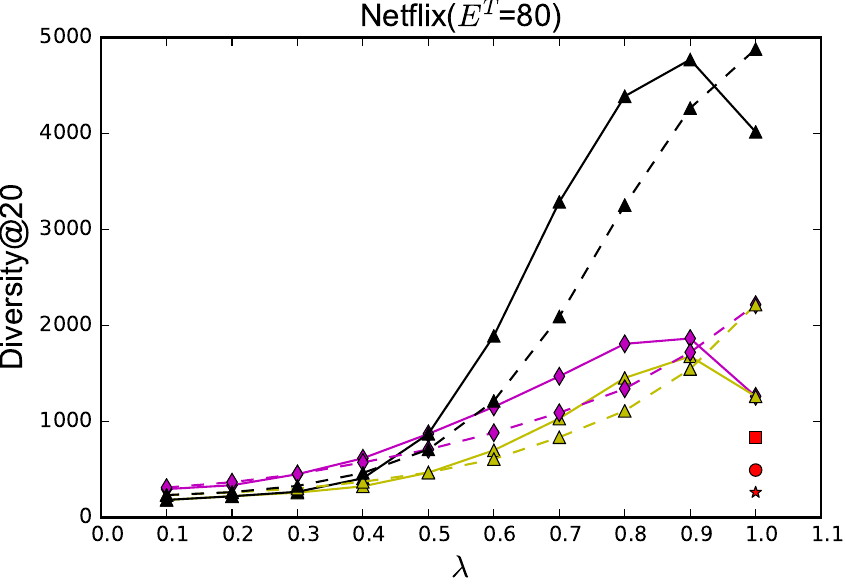}}\\

	\subfloat[]{
		\label{fig:2-6}
		\includegraphics[width=0.7\columnwidth]{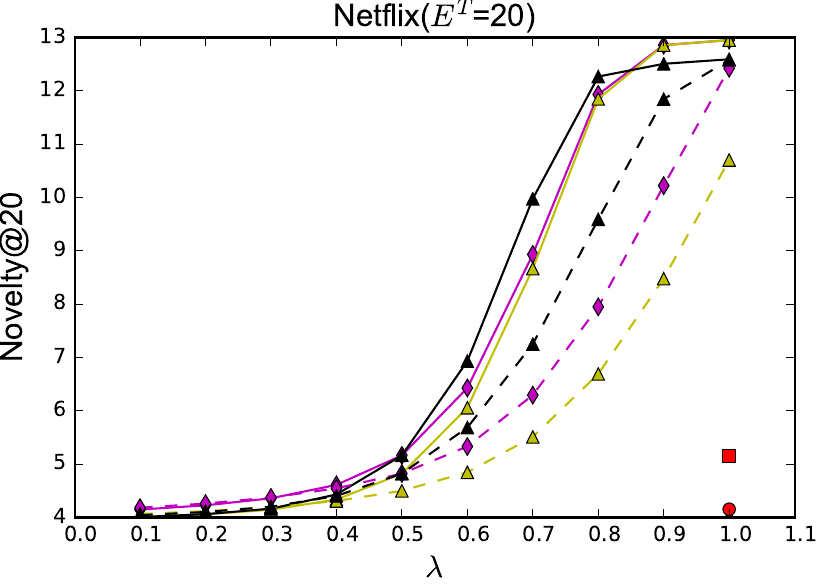}
		\includegraphics[width=0.7\columnwidth]{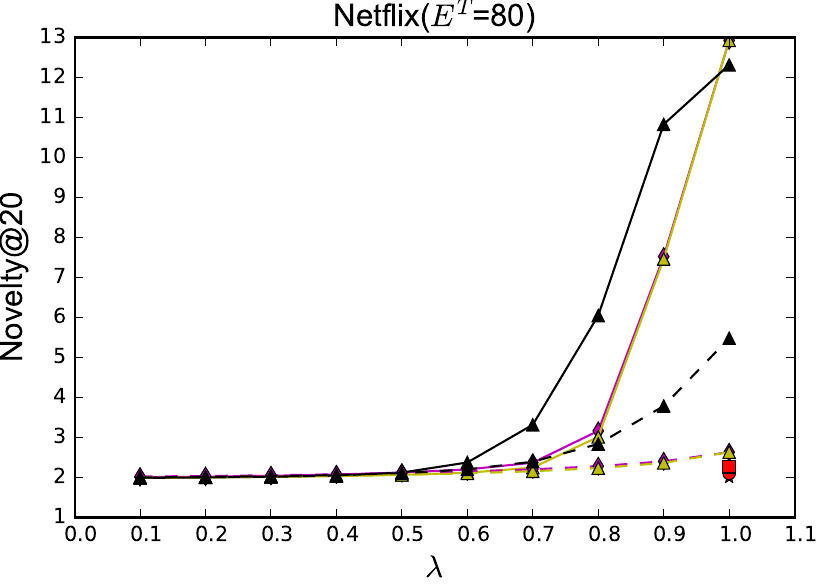}}\\
	
	\captionsetup{justification=centering}
	\caption{Changes of F1-Score@20, Diversity@20 and Novelty@20 with 
 $\lambda$ on Netflix.\\(a) F1-Score@20; (b) Diversity@20; (c) Novelty@20.}
	\label{fig:m-2}
\end{figure*}

{\bf(1)The impact of parameter $\lambda$ on algorithm performance.}

Figures \ref{fig:1-3}, \ref{fig:1-4}, \ref{fig:1-6}, \ref{fig:2-3}, \ref{fig:2-4}, \ref{fig:2-6}, \ref{fig:3-3}, \ref{fig:3-4}, and \ref{fig:3-6} show the variations of F1-Score@20, Diversity-in-top-20, and Novelty@20 of 9 algorithms on two sparsity datasets of MovieLens, Netflix, and RYM datasets at $\epsilon_{opt}$ taken from Table \ref{tab:finalresults-} and Table \ref{tab:final results2} as $\lambda$ changes. It can be seen that:
\begin{figure*}[!h]
	\centering	
	
	\subfloat[]{
		\label{fig:3-3}
		\includegraphics[width=0.7\columnwidth]{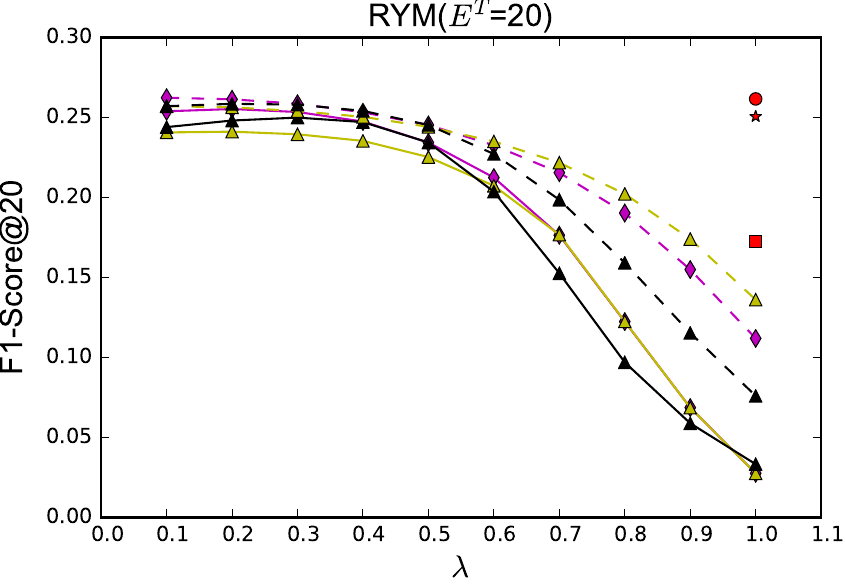}
		\includegraphics[width=0.7\columnwidth]{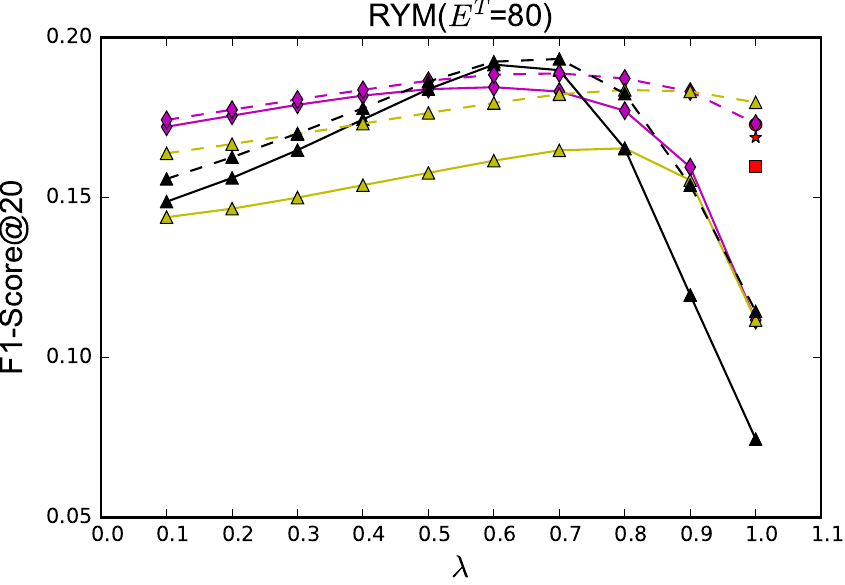}}\\
	
	\subfloat[]{
		\label{fig:3-4}
		\includegraphics[width=0.7\columnwidth]{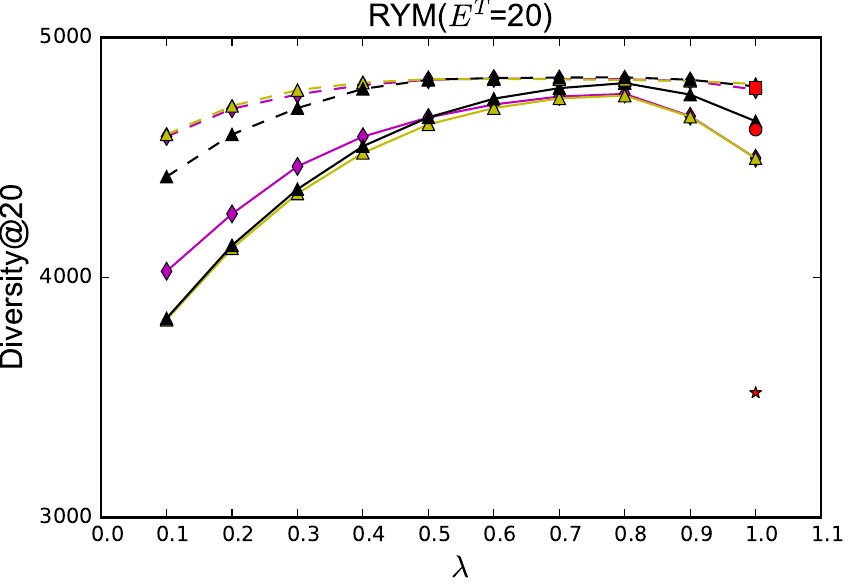}
		\includegraphics[width=0.7\columnwidth]{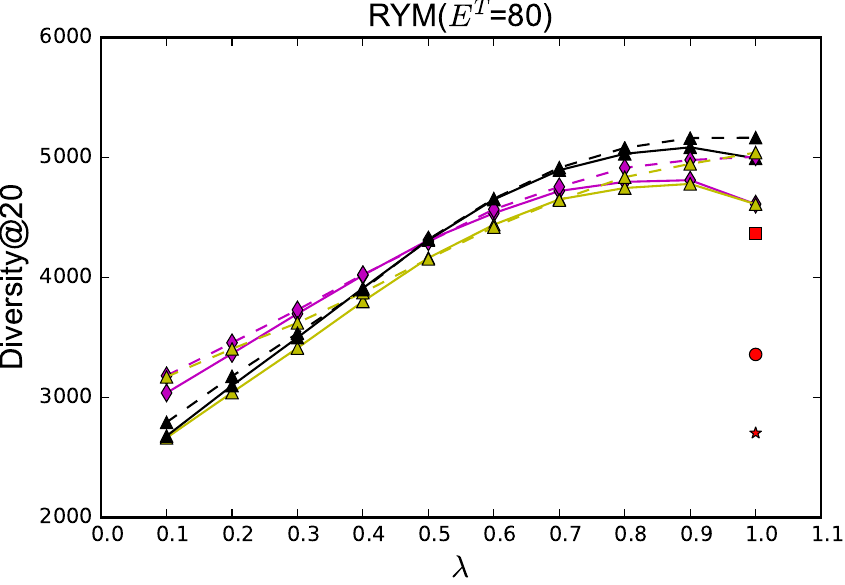}}\\
	
	\subfloat[]{
		\label{fig:3-6}
		\includegraphics[width=0.7\columnwidth]{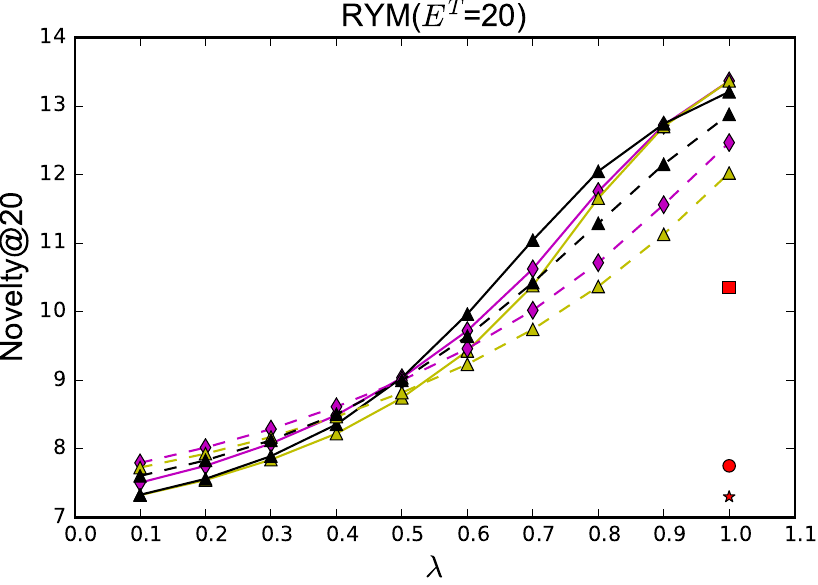}
		\includegraphics[width=0.7\columnwidth]{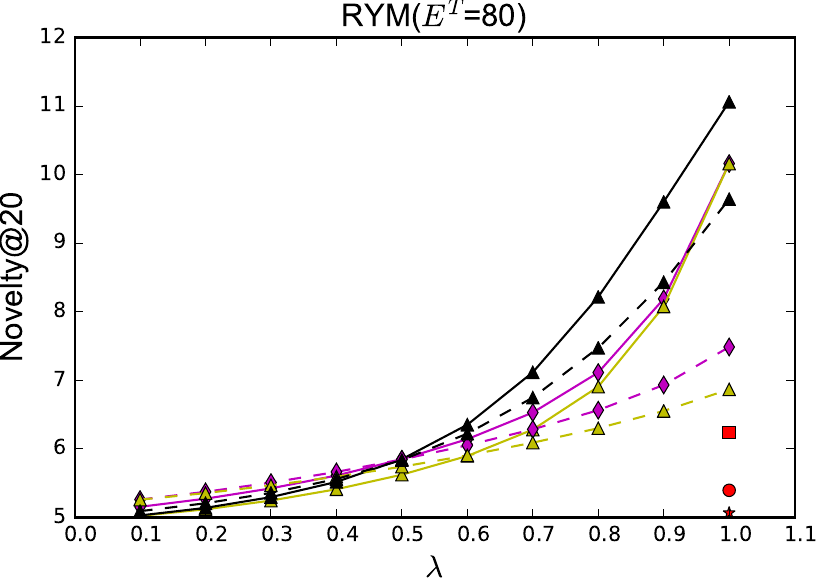}}\\
	
	\captionsetup{justification=centering}
	\caption{Changes of F1-Score@20, Diversity@20 and Novelty@20 with 
 $\lambda$ on RYM.\\(a) F1-Score@20; (b) Diversity@20; (c) Novelty@20.}
	\label{fig:m-3}
\end{figure*}

1. The trends of various algorithms on the MovieLens, Netflix, and RYM datasets are similar. Accuracy gradually decreases as $\lambda$ increases. Diversity initially rises as accuracy decreases and then levels off or slightly decreases.
\begin{figure*}[!h]
	\centering	
	
	\subfloat[]{
		\label{fig:1-1-3}
		\includegraphics[width=0.7\columnwidth]{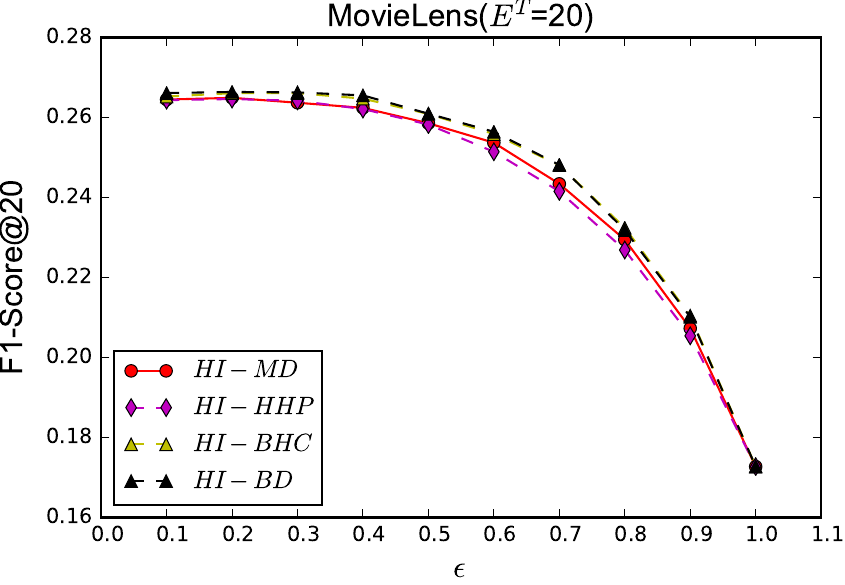}
		\includegraphics[width=0.7\columnwidth]{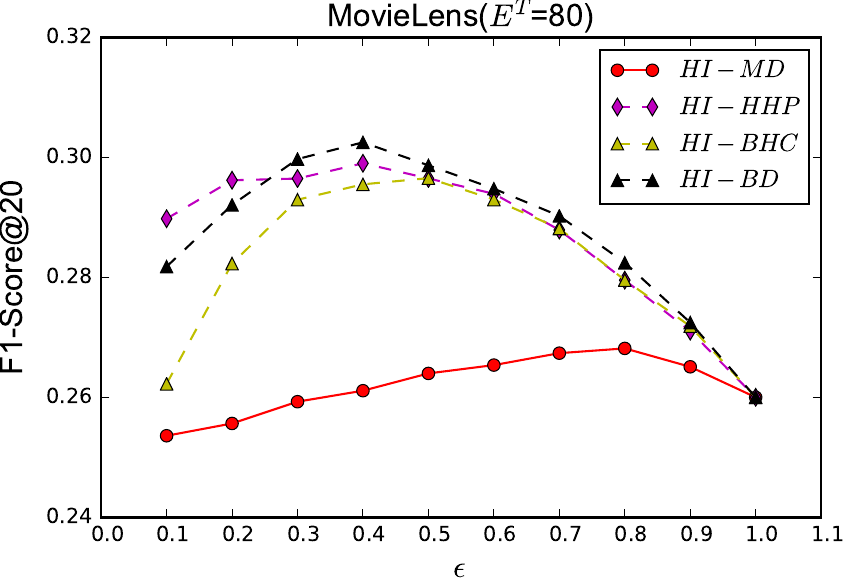}}\\
	
	\subfloat[]{
		\label{fig:1-1-4}
		\includegraphics[width=0.7\columnwidth]{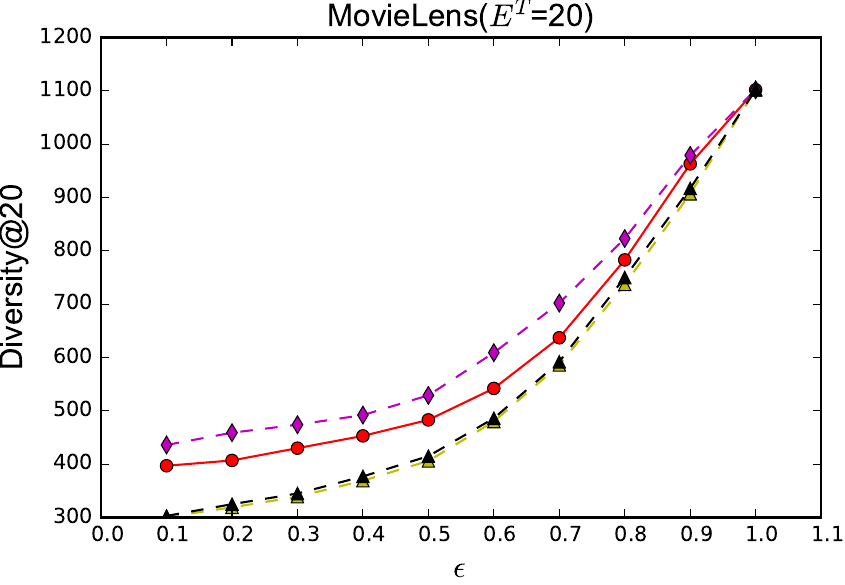}
		\includegraphics[width=0.7\columnwidth]{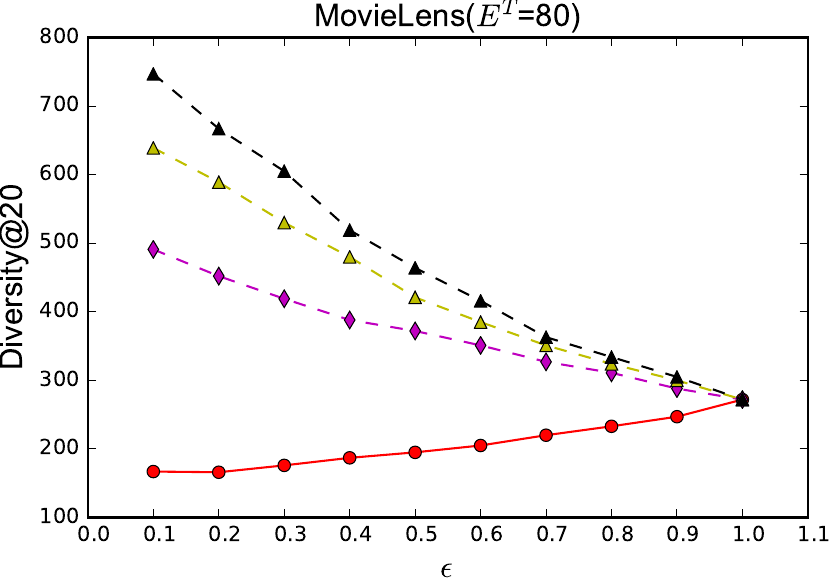}}\\

	\subfloat[]{
		\label{fig:1-1-6}
		\includegraphics[width=0.7\columnwidth]{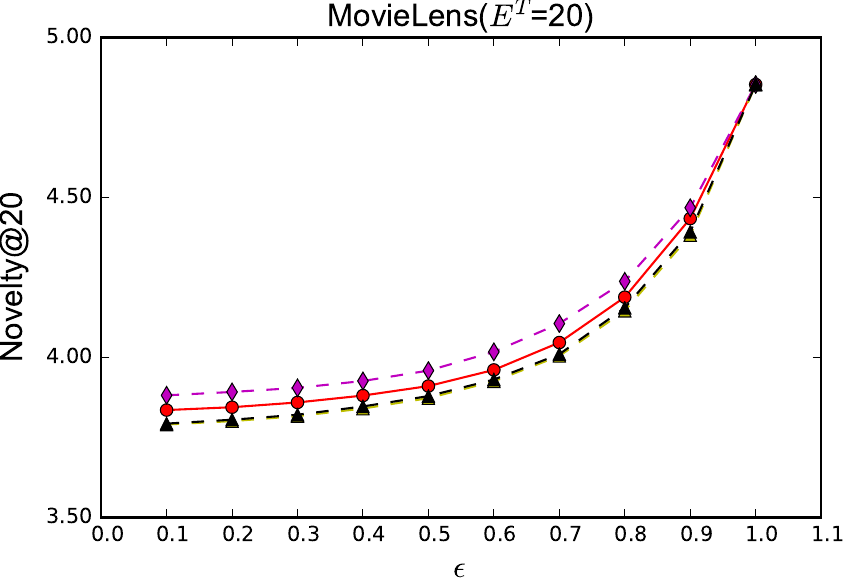}
		\includegraphics[width=0.7\columnwidth]{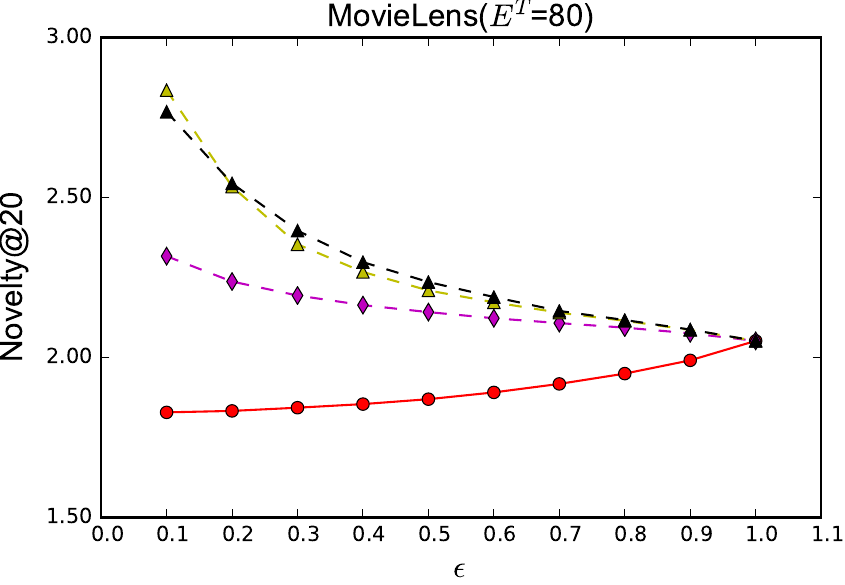}}\\
	\captionsetup{justification=centering}
	\caption{Changes of F1-Score@20, Diversity@20, and Novelty@20 with respect to $\epsilon$ on the MovieLens dataset are as follows: \\(a) F1-Score@20; (b) Diversity@20; (c) Novelty@20.}
	\label{fig:m-4}
\end{figure*}

2. Compared to the original algorithms, the F1-Score and Diversity of the algorithms mixed with ItemCF are generally better. However, in terms of the diversity-Hamming distance HD between user recommendation lists, the hybrid algorithms still show a prominent diversity advantage at smaller $\lambda$ values. But as $\lambda$ increases, the hybrid algorithms are slightly inferior to the original ones, especially on the Netflix dataset.
3. The performance of the hybrid algorithms on datasets with $E^T = 20$ is more prominent than that on datasets with $E^T = 80$. This also shows that the hybrid algorithms perform significantly better on sparse datasets, which is consistent with the previous speculation.

{\bf (2) The impact of parameter $\epsilon$ on algorithm performance.}

\begin{figure*}[!h]
	\centering	
	
	\subfloat[]{
		\label{fig:1-2-3}
		\includegraphics[width=0.7\columnwidth]{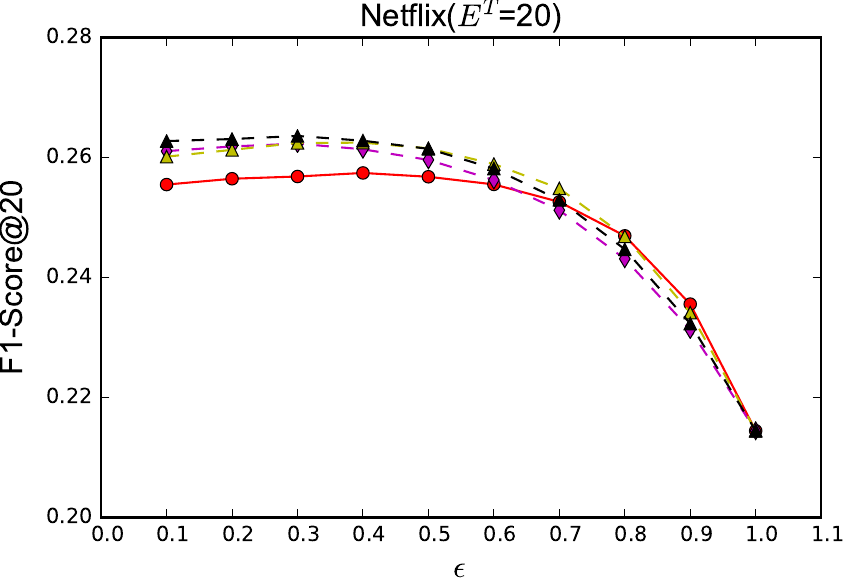}
		\includegraphics[width=0.7\columnwidth]{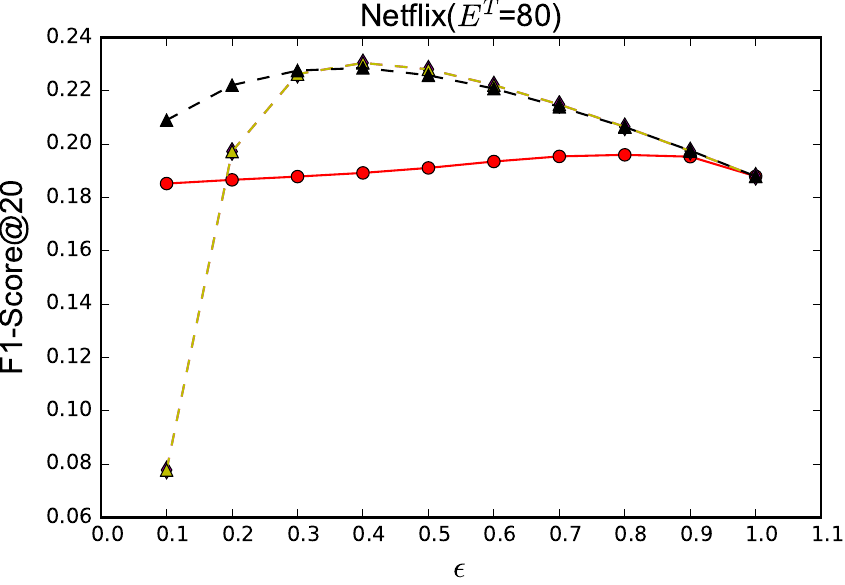}}\\
	\subfloat[]{
		\label{fig:1-2-4}
		\includegraphics[width=0.7\columnwidth]{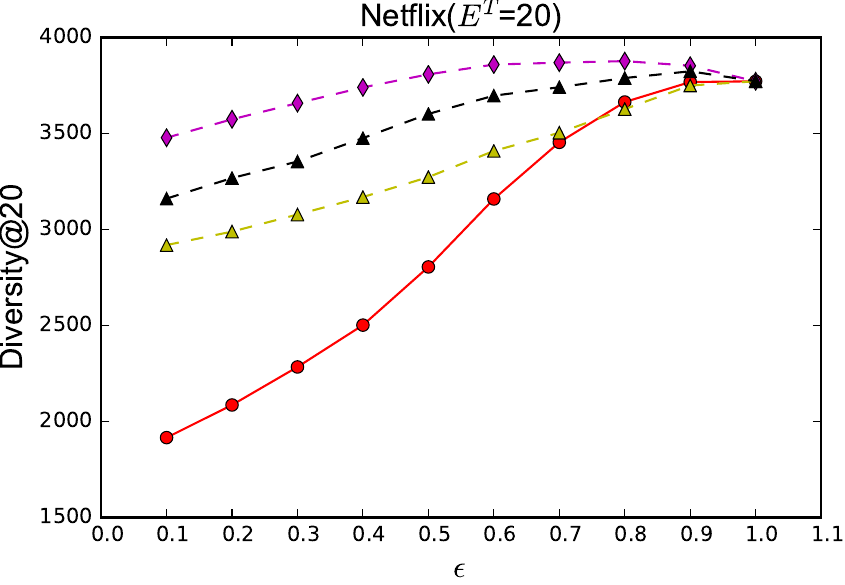}
		\includegraphics[width=0.7\columnwidth]{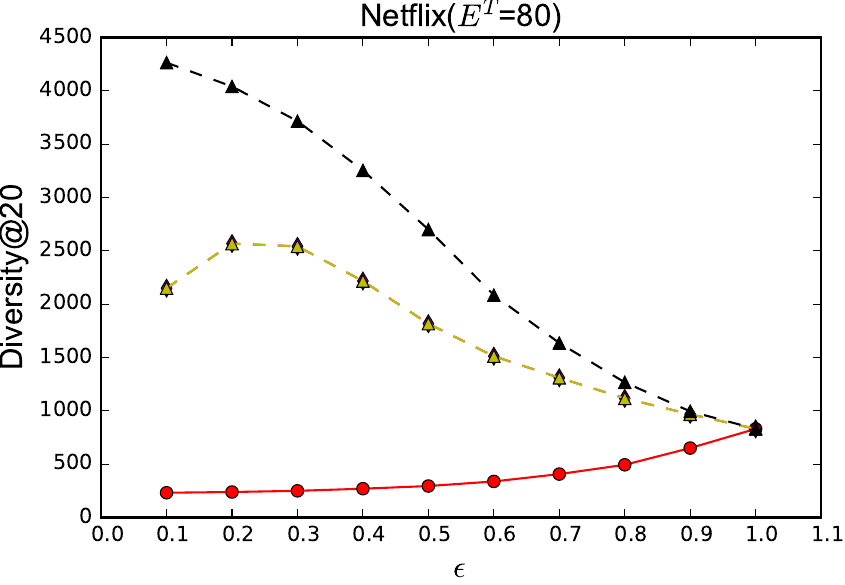}}\\

	
	\subfloat[]{
		\label{fig:1-2-6}
		\includegraphics[width=0.7\columnwidth]{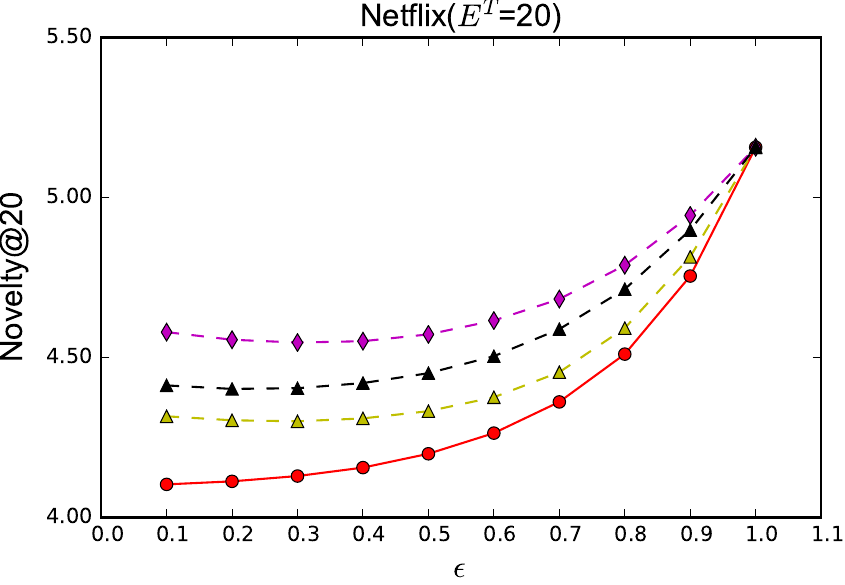}
		\includegraphics[width=0.7\columnwidth]{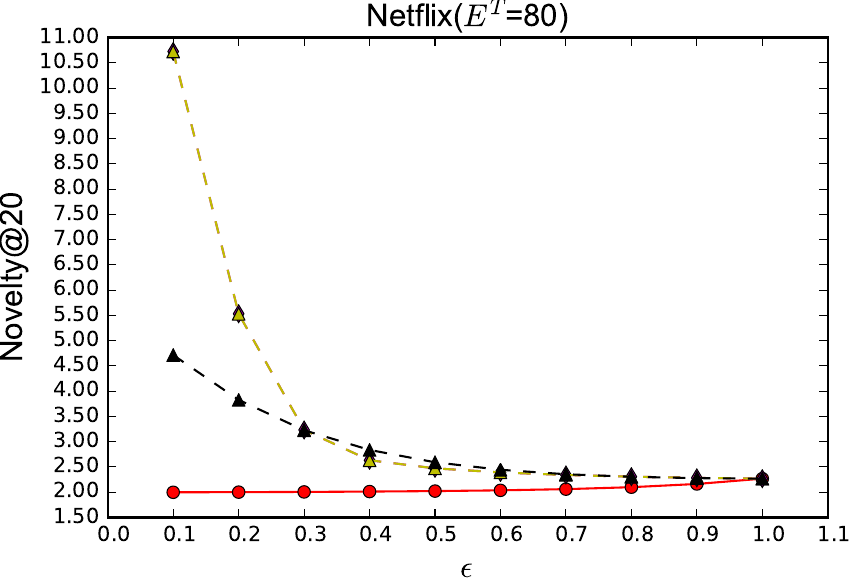}}\\
	\captionsetup{justification=centering}
	\caption{Changes of F1-Score@20, Diversity@20, and Novelty@20 with $\epsilon$ on Netflix. \\(a) F1-Score@20; (b) Diversity@20; (c) Novelty@20}
	\label{fig:m-5}
\end{figure*}
Figures \ref{fig:1-1-3}, \ref{fig:1-1-4}, \ref{fig:1-1-5}, \ref{fig:1-2-3}, \ref{fig:1-2-4}, \ref{fig:1-2-5}, \ref{fig:1-3-3}, \ref{fig:1-3-4}, and \ref{fig:1-3-5} respectively show the variations of F1-Score@20, Diversity-in-top-20, and HD@20 for four hybrid algorithms with $\lambda_{opt}$ taken from Table \ref{tab:finalresults-} and Table \ref{tab:final results2} on three datasets (MovieLens, Netflix, and RYM) and two levels of dataset sparsity. It can be seen that:
\begin{figure*}[!h]
	\centering	
	
	\subfloat[]{
		\label{fig:1-3-3}
		\includegraphics[width=0.7\columnwidth]{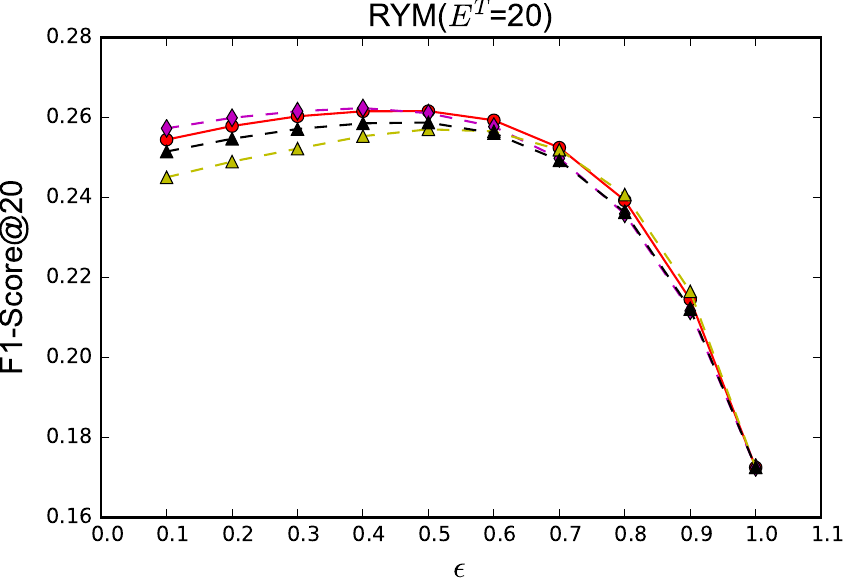}
		\includegraphics[width=0.7\columnwidth]{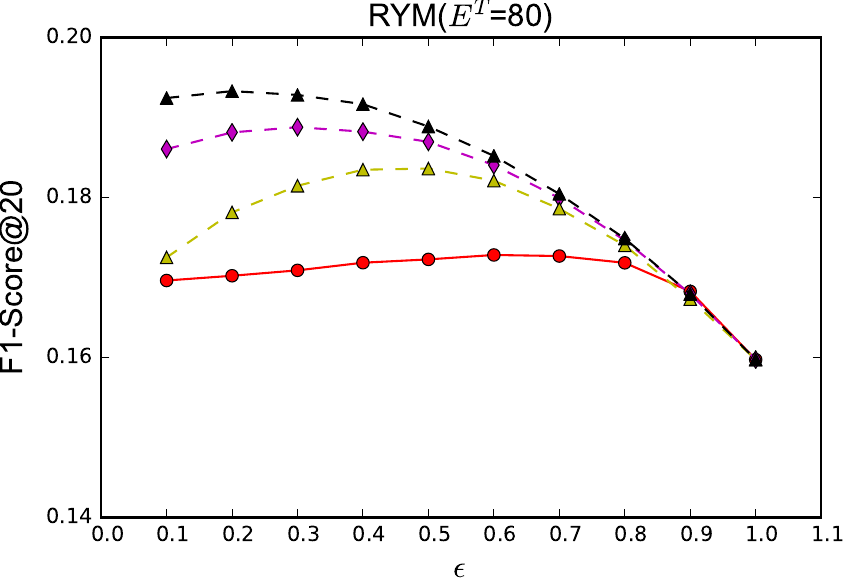}}\\
	
	\subfloat[]{
		\label{fig:1-3-4}
		\includegraphics[width=0.7\columnwidth]{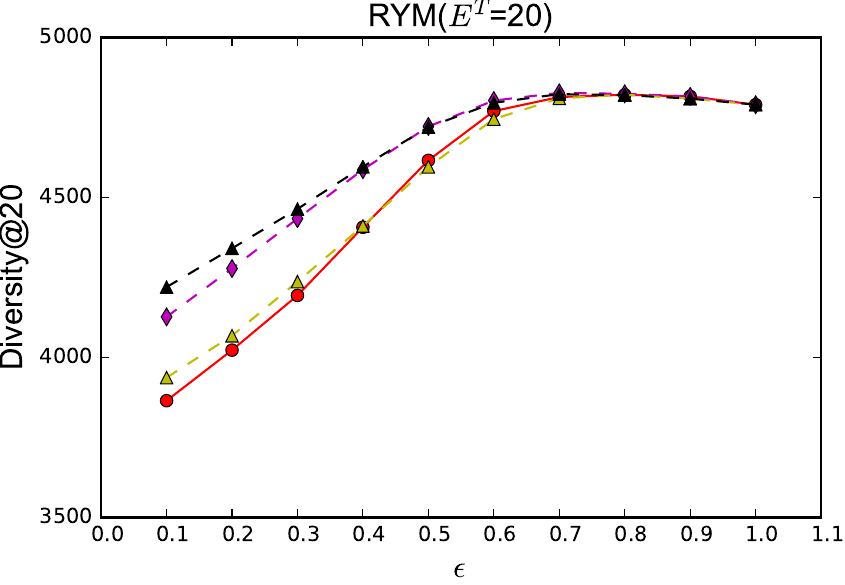}
		\includegraphics[width=0.7\columnwidth]{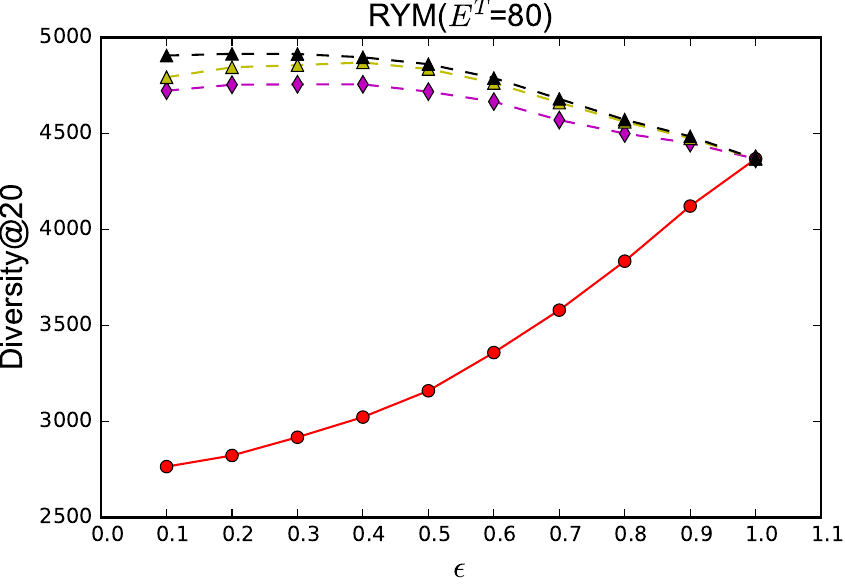}}\\
	
	\subfloat[]{
		\label{fig:1-3-6}
		\includegraphics[width=0.7\columnwidth]{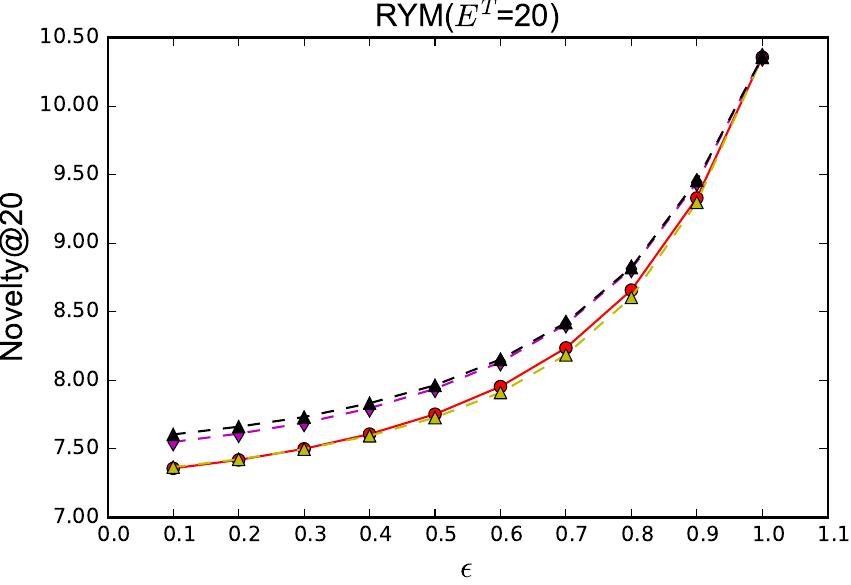}
		\includegraphics[width=0.7\columnwidth]{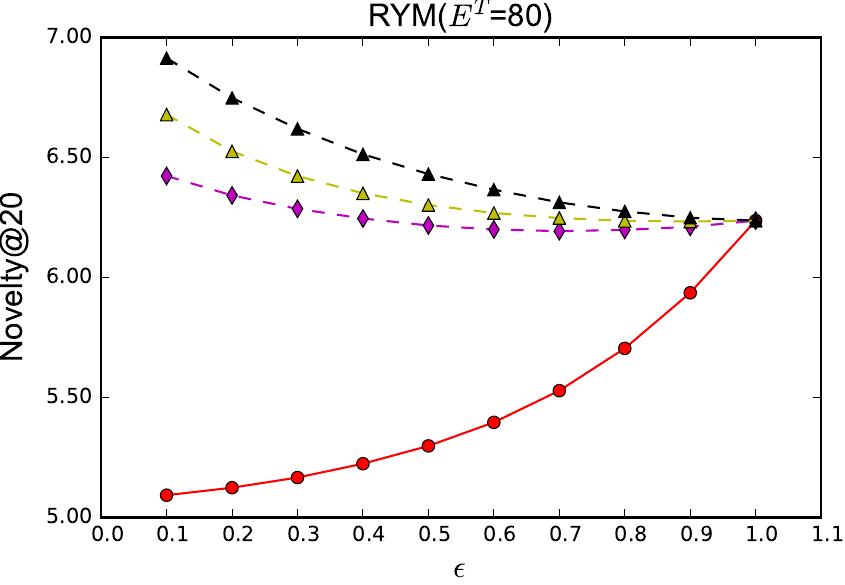}}\\
	\captionsetup{justification=centering}
	\caption{Changes of F1-Score@20, Diversity@20, and Novelty@20 with $\epsilon$ on RYM. \\(a) F1-Score@20; (b) Diversity@20; (c) Novelty@20}
	\label{fig:m-5}
\end{figure*}

1. The trends of the four hybrid algorithms on sparse datasets ($E^T = 20$) are similar. The accuracy decreases steadily as $\epsilon$ increases (from the evolution process of the original MD, HHP, BHC, and BD algorithms to the ItemCF algorithm), while both diversities gradually increase as the accuracy decreases.

2. On dense datasets ($E^T = 80$), among the four hybrid algorithms, HI-BD is significantly higher than the other three algorithms in terms of both accuracy and diversity.

3. On dense datasets ($E^T = 80$), except for HI-MD, the diversity of the other three algorithms decreases monotonically with $\epsilon$, but that of HI-MD increases monotonically. This is mainly because the diversity of the original MD algorithm and ItemCF is much lower than that of the other algorithms, but the diversity of ItemCF is higher than that of MD. Therefore, in the evolution process from the original algorithm to ItemCF, the diversity of MD gradually increases, while that of the other algorithms gradually decreases, and eventually converges to a point, which is the diversity result of ItemCF.

In summary, the hybrid algorithms are better than the original algorithms on both sparse and dense datasets, especially on sparse datasets. The experimental results produced by the mixture of the ItemCF algorithm and other diffusion algorithm types give us more inspiration that making full use of the advantages of various algorithms can achieve the effect of simultaneously improving diversity and accuracy.

\section{Conclusion}
Collaborative filtering algorithms are widely used in the industry. Compared with item-similarity based collaborative filtering algorithms, substance diffusion algorithms have higher accuracy. In this paper, we propose a method that combines item-similarity based collaborative filtering algorithms with substance diffusion. And apply this hybrid idea to the three types of diffusion algorithms (HHP, BHC, and BD) that currently best solve the accuracy and diversity trade-off problem. The experimental results on three real datasets show the effectiveness of the hybrid algorithm. The performance on different sparseness data further demonstrates the robustness of the hybrid algorithm.

Although in general each HI-algorithm mixed with ItemCF performs better in diversity and accuracy than the corresponding original algorithm, the diversity of the ItemCF algorithm has an absolute advantage on some datasets. Therefore, how to improve the accuracy of ItemCF while maintaining diversity is another problem that can be studied during the research process of ItemCF.

\end{document}